\documentclass[structabstract]{aa}  
\usepackage{graphicx}
\usepackage[dvips]{color}
\usepackage{txfonts}
\usepackage{natbib}
\begin{document}
   \title{Angular momentum transfer between oscillations and rotation in subdwarf B hybrid pulsators}

   \author{F. P\'erez Hern\'andez
          \inst{1,2}
          \and
	  R. Oreiro\inst{3}
	  \and 
	  Haili Hu\inst{4}
          }

   \institute{Instituto de Astrof\'\i sica de Canarias (IAC), E-38200 La Laguna, Tenerife, Spain
         \and
              Departamento de Astrof\'\i sica, Universidad de La Laguna (ULL), E-38205 La Laguna, Tenerife, Spain\\
              \email{fph@iac.es}
         \and
              Instituto de Astrof\'\i sica de Andaluc\'\i a, IAA (CSIC), Glorieta de la Astronom\'\i a, s/n E-18008 Granada, Spain\\ \email{roreiro@iaa.es}
         \and 
         Institute of Astronomy, The Observatories, Madingley Road, Cambridge CB3 0HA \\ 
               \email{hailihu@ast.cam.ac.uk}
             }

   \date{Received xx, 2011; accepted xx, 2011}

 
  \abstract
    {Subdwarf B pulsators exhibit pressure ($p$) and/or gravity ($g$) modes. Their frequency spectra range from very simple, with
few frequencies, to very rich, with more than fifty peaks in some cases. Balloon09 is a hybrid pulsating subdwarf B,
showing a large number of $p$- and $g$-modes including a triplet and a quintuplet-like structures which are interpreted as 
$p$-mode frequency splittings due to stellar rotation. Photometric observations undertaken in two subsequent years revealed 
a change in these rotational splittings of $0.24-0.58\,\mu$Hz/yr.}
    {We analyse the possibility of angular momentum interchange between stellar rotation and internal gravity waves as a 
mechanism for the observed rotational splitting variations.}
   {The expected change in the rotational splitting of eigenmodes resulting from this mechanism are computed in the 
non-adiabatic linear approximation for a stellar structure model that is representative of the target star. 
The fact that $g$-modes are also observed in Balloon09 makes it a particularly suitable candidate for our study, because 
the change in the rotational splittings are proportional to the amplitude squared of the $g$-modes which, in this case, 
can be estimated from the observations.}
   {We find that this mechanism is able to predict changes in the splittings that can be of the same order of magnitude 
as the observed variations.}
   {}


   \maketitle
%

\section{Introduction}

Subdwarf B (sdB) stars populate the blue extension of the horizontal branch (EHB) in a Hertzsprung-Russell diagram 
\citep{heber86}. Their high
temperatures ($T_{\rm eff}>20\,000\,$K) and high gravities ($5<\log g<6$) correspond to an evolutionary state in between
the red giant and white dwarf phases, although the detailed history is not well understood yet. An anomalously high 
mass loss rate at some point is needed to explain the hot subdwarf's thin H-envelope, which prevents them from
ascending the asymptotic giant branch \citep{dorman93,dcruz96}. Mass transfer due to binary interaction has been invoked as
an explanation, as it is observationally supported by the high binary fraction of these objects in the field.  
The predominantly single sdBs that populate the EHB in globular clusters \citep{moni11} can be explained by white dwarf 
mergers (Han et al. 2008).

Some sdB stars exhibit stellar oscillations. Indeed, long- ($\sim$1\,h) and short-period ($\sim$10\,min)
pulsating sdBs can be distinguished \citep{kilkenny97,green03}, with a few hybrid objects showing both period regimes
\citep{schuh06,oreiro05,lutz09,baranrat}. Long-period oscillations are explained in terms of gravity modes 
\citep{fontaine03}, 
while short-period variations are attributed to pressure modes \citep{charpi97}. Asteroseismic techniques
can thus be applied to sdBs to retrieve information on their internal structure and rotation, and 
will eventually help to constrain their evolutionary formation channels~\citep{hu08}.

Balloon\,090100001 (Bal09 hereafter) is a hybrid sdB pulsator, with a rich frequency spectrum both at low and high 
frequencies. Together with the facts that it is the brightest sdB pulsator and has one of the highest amplitude of 
oscillation, make Bal09 a very interesting object.
Since the discovery of its pulsating nature \citep{oreiro04}, a multi-site photometric campaign was organized to unravel 
the complex frequency structure of the
target during the 2005 summer \citep{baran09}, while a low resolution time-resolved spectroscopy \citep{telting06} and a 
high resolution spectroscopy dataset enabled a mode identification \citep{telting08,baran08}. 
Mode identification was also attempted through multicolour photometry \citep{charpi08}, 
while \cite{vangrootel08} provide a seismic solution for Bal09.

An interesting feature concerning Bal09's amplitude spectrum is that a triplet-like structure is resolved close to the
dominant, single mode. Moreover, five components are also identified $\sim$25\,$\mu$Hz apart from the triplet. 
From these modes a frequency splitting of $\sim 1.5\,\mu$Hz is derived. Assuming a canonical total mass 
$M=0.5M_{\odot}$ and using the spectroscopic $\log g=5.39$ value, a rotational velocity
between 1.5 -- 2\,km/s \citep{telting08} is obtained, in
agreement with the typical low surface-rotation speed of single sdBs ~\citep{geier10}.

Bal09 has been photometrically monitored after the main multi-site campaign in 2005 
but no result has been published yet (Baran, private communication). A comparison
of the 2004 and 2005 seasons, allowed the detection of changes in the amplitudes of many modes.
Amplitude variations are not uncommon among pulsating sdBs, as it is discussed in \cite{kilkenny10}. 
On the other hand, small changes in the frequency splitting are found from one campaign to the other as well.
Although close to the frequency
resolution, the triplet components have moved in the right direction as if rotation is varying. The quintuplet also
varies in the same way, despite the fact that not all five components were resolved in 2004. 
Since it is unusual for a star to change its rotation in such a short time scale, a physical explanation for this 
phenomenon is yet to be found.

The aim of this work is to examine one possible mechanism that could explain the changes in the frequency splitting. We 
analyse here the possibility of angular momentum interchange between stellar rotation and internal waves. 
This scenario requires the existence of internal gravity waves with a dissipation mechanism or any other 
non-conservative process that eventually could lead to amplitude variations. This is the case in Bal09, 
where we can estimate 
the $g$-mode energies from the observations and hence quantify the angular momentum transport by waves.

The possibility of angular momentum interchange between waves and rotation
was originally proposed by~\cite{ando81,ando83}, where it is described how non-axisymmetric, non-radial 
oscillations can redistribute angular momentum in stars. The complete formalism is included in these works, 
and is also applied to two $\beta$ Ceph stars. He concluded that non-radial oscillations can affect significantly the 
rotation profile of these stars, but in a time scale ($10^4$\,yr) much larger than the one observed in our case.
A related process has been postulated for the Sun~\citep{zahn97,kumar99,talon02}
to explain the flat Sun's rotation profile in the radiative interior. In that case however, the angular momentum is carried 
out by very low frequency gravity waves generated at the base of the convection zone, which are then completely 
damped in the radiative interior.
The time scale for such a momentum transfer is about $10^7$\,yr for solar-like stars~\citep{talon02}.
 
\section{Observational data}

\begin{figure}
\centering
\includegraphics[scale=0.7]{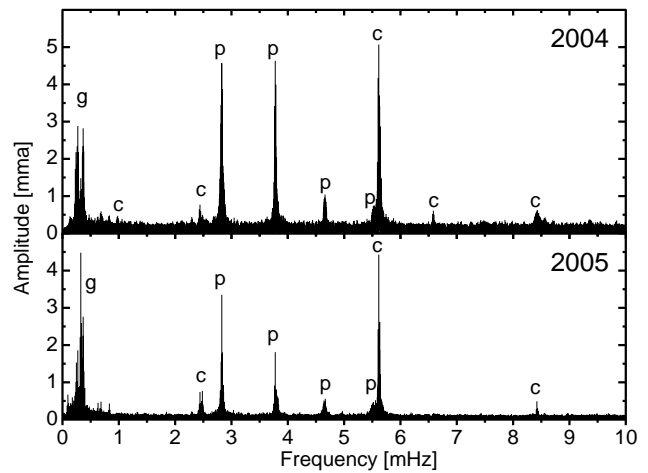}
      \caption{Comparison of the amplitude spectrum of Bal09 in 2004 and 2005 after removing the 3 highest peaks.
Figure from \citet{baran09}. $P$, $g$, and $c$ stand for $p$-modes, $g$-modes, and combination
frequencies respectively.}
         \label{fig:20042005}
\end{figure}

In Fig.~\ref{fig:20042005} we compare the amplitude spectrum of the target obtained from the 2004 and 2005 
photometric campaigns. The figure shows the residual spectra after the three highest amplitude $p$-modes
(at $\sim$2.8\,mHz) have been prewhitened. In general, lower amplitudes are obtained for $p$-modes in 2005, 
whereas similar or larger are obtained for $g$-modes. 
This suggests that $p$- and $g$-modes are interchanging energy, a fact that could be related to the changes in the
rotational splittings, but in any case the analysis of this process is outside the scope of the present work.
The highest amplitude $g$-modes for the two photometric runs are compared in Table~\ref{tbl:baran08}.
A complete comparison of modes in the two sessions is found in \cite{baran09}.

\begin{table}
\caption{Frequency analysis for the highest amplitude $g$-modes based on photometric data.
From \citet{baran09}. }
\label{tbl:baran08}      
\centering                          
\begin{tabular}{c c| c c }        
\hline\hline                 
\multicolumn{2}{c|}{2004} & \multicolumn{2}{c}{2005}\\
\hline
Freq. & Ampl.  & Freq. & Ampl. \\    
($\mu$Hz) & (mma)  & ($\mu$Hz) & (mma)  \\    
\hline        
229.57 & 0.57 & 229.55 & 0.85\\
239.97 & 1.96 & 239.97 & 1.08\\
246.31 & 0.61 & 246.30 & 1.52\\
272.38 & 2.75 & 272.46 & 1.72\\
298.89 & 0.67 & - & - \\
325.67 & 1.25 & 325.61 & 4.44\\
331.21 & 0.81 & 331.18 & 0.75\\
365.81 & 2.66 & 365.81 & 2.65\\
397.19 & 0.49 & 397.23 & 0.84\\
631.08 & 0.44 & 630.74 & 0.42\\
684.35 & 0.60 & 684.40 & 0.47\\
833.08 & 0.53 & 833.09 & 0.43\\

\hline                                   
\end{tabular}
\end{table}

\begin{figure}
\centering
\includegraphics[]{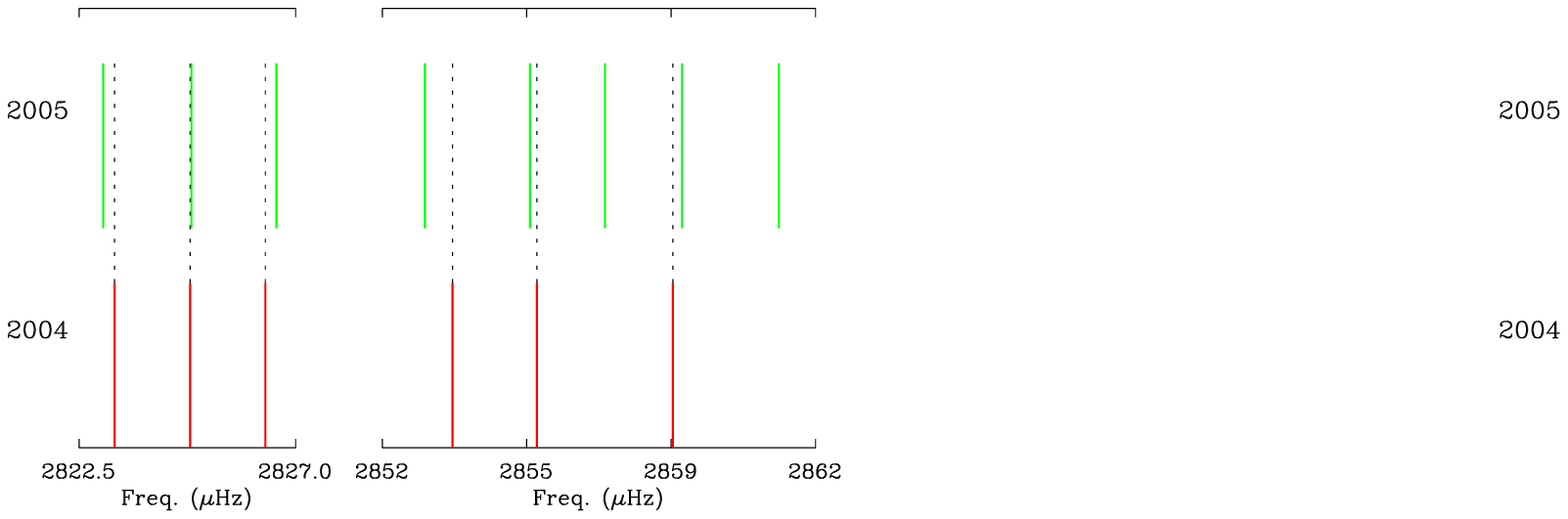}
   \caption{Splittings of the $\ell=1,\nu=2807.5\,\mu$Hz (left panel) and $\ell=2,\, \nu=2853.4\,\mu$Hz (right panel) 
modes in 2004 and 2005. From \citet{baran09}.} 
      \label{fig:split}
\end{figure}

In Fig.~\ref{fig:split} a closer view to the the triplet and quintuplet structures is shown as observed in 2004 and 2005.
Interestingly, these changes do not seem random, but as 
if the mean rotational velocity of the star was increasing in
the time scale of the observations. The retrograde $|m|=1$ triplet component changed by $0.24\,\mu$Hz/yr from 2004 
to 2005, while the retrograde $|m|=2$ quintuplet component changed its frequency as much as $0.58\,\mu$Hz/yr in the same 
period.

Telting \& \O stensen (2006) publish frequencies and line-of-sight velocities based on low resolution mode spectroscopy
acquired during the summer of 2004. Their results are included in Table~\ref{tbl:telting06}.
High-resolution spectroscopy during the summer 2006 was obtained by~\cite{telting08}. They only list the line-of-sight 
velocity for the fundamental mode: 14.5 km/s, which is significantly lower than 18.9\,km/s measured in 2004.
In both cases, line-of-sight velocities were computed from cross-correlation
profiles (ccp), which merge all the spectral lines into a single, and with higher signal-to-noise artificial line. Thus, 
spectral features formed at different depths are combined to produce the ccp's. Moreover, Balmer lines are used in the 
low-resolution dataset by~\cite{telting06}, while only metallic ones are cross-correlated from the high-resolution data 
of~\cite{telting08}. Hence, in this case, the change in amplitude can be attributed to the different techniques used.

\begin{table}
\caption{Frequency analysis based on low-resolution spectroscopy by Telting \& \O stensen (2006).
The observations were carried out in 2004.}  
\label{tbl:telting06}      
\centering                          
\begin{tabular}{c c }        
\hline\hline                 
Frequency & RV \\    
($\mu$Hz) & (km/s) \\    
\hline        
272.44 & 0.86\\
325.67 & 0.96 \\
365.77 & 1.05 \\
2807.47 & 18.89 \\
2823.24 & 5.88 \\
2824.81 & 3.44 \\
2826.28 & 1.86 \\               
\hline                                   
\end{tabular}
\end{table}

\section{Structure models and non-adiabatic oscillations}

\subsection{Model 1}

For the most part of the work we have considered a single stellar structure model that closely matches the fundamental 
parameters of Bal09 (Model 1 hereafter). 
It is a $0.43 M_\odot$ total mass model with a hydrogen amount of $2\times10^{-4}  M_\odot$, $T_{\rm eff}=26\,800$\,K, 
and $\log g=5.47$. Note that the stellar mass equals the seismic mass  found by \citet{vangrootel08}. 
The $T_{\rm eff}$ and $\log g$ are within the observed spectroscopic errors considered by these authors, 
although we chose $T_{\rm eff}$ on the low side to ensure unstable $g$-modes.

The structure model is constructed with a version of the stellar evolution code STARS \citep{Eggleton71},
updated for asteroseismology. We started the evolution at the ZAEHB, and the particular model we consider is near the end 
of core He burning with an EHB age of $2.1\times10^8$ yr. H/He diffusion is included consistently during the evolution, 
whereas the effects of diffusion of Fe and Ni are approximated with a Gaussian accumulation around $\log T=5.3$. 
Radiative opacities that account for the Fe/Ni-enhancement are taken 
from the Opacity Project \citep{Badnell05}.
More details and input physics are given in \citet{Hu09} and references therein. 

Normal modes of oscillation and eigenfunctions for $\ell \leq 6$ were computed in the linear, non-adiabatic approximation 
with the pulsation code MAD \citep{Dupret01}.
The structure model was chosen so that the fundamental mode ($\nu=2.786\,$mHz) is close to Balloon's observed
highest amplitude mode at $\nu =2.807\,$mHz. This frequency does not show any rotational splitting
and has been identified as a radial mode~\citep{baran08,telting08,charpi08}.

\begin{figure}
\centering
\includegraphics{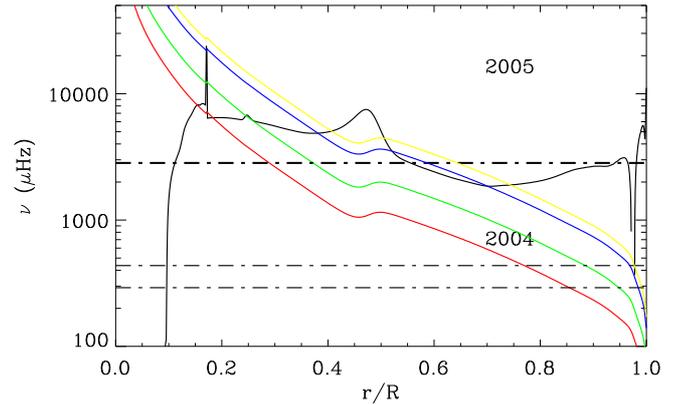}
\caption{The black solid line is the Brunt-V\"ais\"al\"a frequency, $N$, while red, green, blue, and yellow lines are the 
Lamb acoustic frequencies, $S_{\ell}$, for $\ell=1$, $2$, $4$, and $5$, respectively.
The dashed horizontal lines correspond to the modes indicated in the text (the $\ell=1$ and $\ell=2$ $p$-modes
look like a single thick line). Model 1 was used.} 
\label{fig:n2s2}
\end{figure}

In Fig.~\ref{fig:n2s2} the Brunt-V\"ais\"al\"a ($N$) and Lamb ($S_{\ell}$) frequencies for this model are
represented. They are displayed in terms of cyclic frequencies so that the propagation cavities are easily delimited in 
comparison with observed modes. Peaks in $N$ denote gradients in the composition
profile, hence, we can distinguish the H envelope 
(from $r/R\sim 0.45$ to the surface), the He radiative layer ($0.2\leq r/R \leq 0.45$) and the inner 
convective CO core formed as 
ashes of the He-core burning process. The horizontal lines in the figure correspond to the frequencies of some selected 
modes: two $p$-modes at $\nu=2807.5$ $\mu$Hz and $\nu=2853.4$ $\mu$Hz, which are the $\ell=1$ and $\ell=2$ theoretical 
counterparts of the observed splittings, and two low frequency $g$-modes: $\ell=4$, $\nu=436.0\,\mu$Hz
and $\ell=5$, $\nu=291.4\,\mu$Hz.

\begin{figure}
\centering
\includegraphics{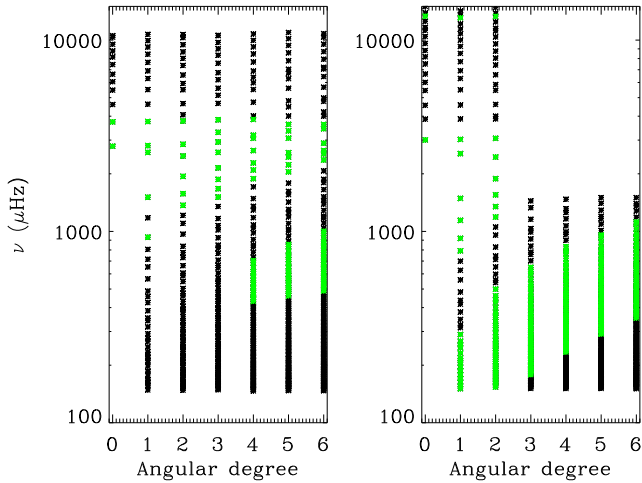}
\caption{Theoretical eigenfrequencies as function of angular degree. Modes with angular degrees $0\leq \ell \leq 6$ and 
cyclic frequencies $\nu>150\,\mu$Hz were considered. Excited modes are shown in green. Left panel is for Model 1 and right 
panel for Model 2.}
\label{fig:growth}
\end{figure}

In the left panel of Fig.~\ref{fig:growth} we show all the computed eigenfrequencies for this structure model, indicating 
if the mode is linearly unstable (in green) or not (in black). 
Low order $p$-modes ($2800<\nu\lesssim4000$ $\mu$Hz) are excited for all the angular
degrees ($\ell$) considered, which can account for the high-frequency peaks detected in Bal09. Another island of
unstable modes is due to high order $\ell>3$ gravity modes. Their frequencies, within $400-900$ $\mu$Hz are slightly
shifted compared to the observed low-frequency peaks in Bal09. The relevance of this fact in our work will be discussed
later on.

\begin{figure}
\centering
\includegraphics{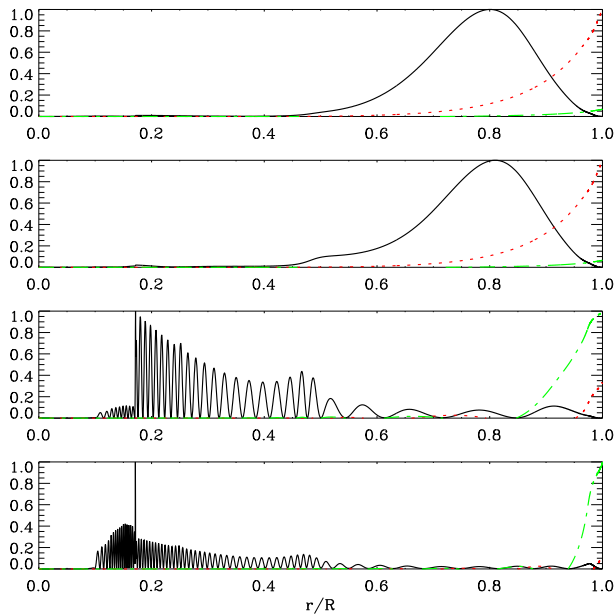}
\caption{Radial (red, dotted line) and horizontal (green, dot-dashed line) components of the eigenfunctions normalized
such that $\xi_r(R)=1$ ($p$-modes) or $\xi_{\rm h}(R)=1$ ($g$-modes), where $R$ is the photospheric radius, for some 
oscillation modes: From top to bottom: $\ell=1$, $\nu=2807.5\,\mu$Hz, $\ell=2$, $\nu=2853.4\,\mu$Hz,
$\ell=4$, $\nu=436.0\,\mu$Hz and $\ell=5$, $\nu=291.4\,\mu$Hz. Also shown (black solid line) is the integrand in the 
dimensionless kinetic energy ${\cal E}$ defined by Eq. \ref{eq:aener}. Model 1 was used.}
 \label{fig:autof}
\end{figure}

The radial and horizontal components of the displacement eigenfunction vector for some particular modes (those indicated in 
Fig.~\ref{fig:n2s2}) are included in Fig.~\ref{fig:autof}. 
The contribution of every point in the star to the dimensionless kinetic energy, ${\cal E}$, given by
\begin{equation}
{\cal{E}} = \frac{4 \pi \int_0^R \rho r^2 \left[\ell(\ell+1)|\xi_h|^2+ |\xi_r|^2\right] dr}
{M \left[\ell(\ell+1)|\xi_h(R)|^2+ |\xi_r(R)|^2\right]}
\; ,
\label{eq:aener}
\end{equation}
is also included in Fig~\ref{fig:autof}.
The first two modes on the top are the theoretical eigenfrequencies representative of the observed triplet 
and quintuplet respectively. 
The $\ell=4$ and the $\ell=5$ are in the observed $g$-mode region, albeit only the $\ell=4$ is theoretically unstable. 

\subsection{Model 2}

One of the problems with the current sdB models is that the observed instability strip of the $g$-modes can not be correctly 
modeled.
Compared to the observations, typically, the $T_{\rm eff}$'s of models with unstable $g$-modes are too low, the periods of 
unstable $g$-modes are too low and only modes with high spherical degree ($\ell\geq 3$) are excited 
\citep[see e.g][]{fontaine03,Jeffery06,Hu09}. This problem is also apparent in our Model 1 as 
discussed in the previous section. However, Hu et al. (in prep.) have computed new models with atomic diffusion, including 
radiative levitation of H, He, C, N, O, Ne, Mg, Fe and Ni, that solve the aforementioned issues with the unstable $g$-modes. 
Hence, we also compare with such a preliminary model (Model 2 hereafter). Model 2 is obtained from the same ZAEHB model as 
used for Model 1 but now the evolution and pulsations are computed with fully self-consistent atomic diffusion.

The right panel of Fig.\ref{fig:growth} shows the stable and unstable modes for Model 2. 
In this case  there are a large number of unstable modes in the observed frequency range, between $200\,\mu$Hz and
$800\,\mu$Hz and including degrees from $\ell=1$ to 6. Note that for higher degrees the frequency range of unstable
modes is shifted upwards.

\section{Transfer of angular momentum and change in the rotational splitting}

Non-radial, non-adiabatic oscillations, especially $g$-modes with large tangential velocities, can transfer angular 
momentum to the mean rotational flow. This variation in the rotational velocity will consequently affect the frequency 
splitting of non-radial $g$ and $p$-modes.

\subsection{Transfer of angular momentum due to gravity modes}

The angular momentum transfer due to nonaxisymmetric waves in stars can be described in terms of a Reynolds 
stress~\citep{ando81}, namely:
\begin{equation}\label{eq:tau}
\frac{\partial}{\partial t} (\rho \varpi^2 \Omega) = -\tau
\quad \mbox{where} \quad
\tau \equiv \nabla_{\ell}\cdot(\rho \varpi\overline{\vec{V}_p V_{\phi}})
\; .
\label{eq:changej}
\end{equation}
Here $\varpi$, $\vec{V}_p$, $V_{\phi}$ and $\nabla_{\ell}$ are the cylindrical polar radius, poloidal
component of the perturbed velocity, azimuthal component of the perturbed velocity, and poloidal component of the 
differential operator $\nabla$, respectively. The rotation angular velocity is given by $\Omega$. Here and in the following 
we will use a similar notation to \cite{unno89}. 

Taking the linear and Cowling approximations in the wave equations, assuming a dependence of the form 
$\exp{[i(\sigma t + m \phi)]}$, and if the star rotates slowly, such that the real part of the eigenfrequencies 
obey $\sigma_{\rm R} \gg \Omega$, the change in the angular momentum can be expressed as:
\begin{equation}
\tau= \tau_{\rm wave}+\tau_{\rm NA}
\label{eq:taus}
\end{equation}
where the terms $\tau_{\rm wave}$ and $\tau_{\rm NA}$ are given by: 
\begin{equation}\label{eq:tauwave}
\tau_{\rm wave}=\sigma_{·\rm R}\sigma_{\rm I} m\rho\biggl[\frac{\ell(\ell+1)|\sigma|^2}{S_{\ell}^2}|\xi_h|^2+
  \frac{N^2}{|\sigma|^2}|\xi_r|^2\biggr]|Y_{\ell}^m|^2
\label{eq_tauwave}
\end{equation}
\begin{equation}\label{eq:tauna}
 \tau_{\rm NA}= \frac{1}{2} m P\, {\rm Im} \biggl[\biggl(\frac{\partial P}{P}\biggr)^*\biggl(\frac{\partial
\rho}{\rho}\biggr)\biggr]|Y_{\ell}^m|^2
\; .
\label{eq_tauna}
\end{equation}
Here $\sigma=\sigma_{\rm R}+i \sigma_{\rm I}$ is the complex eigenfrequency of the mode and 
$Y_{\ell}^m$ the spherical harmonic.  
$\tau_{\rm NA}$ represents the transfer of angular momentum due to non-adiabatic effects. In this approximation 
$\tau_{\rm NA}$ is equal to the work function, $w(r)$, except for a constant factor.
On the other hand, the term $\tau_{\rm wave}$ represents a wave transience.
As explained in \cite{ando81}, in Eq. \ref{eq_tauwave} for $\tau_{\rm wave}$, the imaginary part of the
eigenfrequency, $\sigma_{\rm I}$, comes from a time derivative of the wave amplitude and hence can be derived 
appropriately. In particular, it makes sense to take $\sigma_{\rm I}$ from the observed changes in the amplitudes of 
the $g$-modes, provided they are intrinsic and not caused by beating phenomena.

\begin{figure}
\centering
\includegraphics{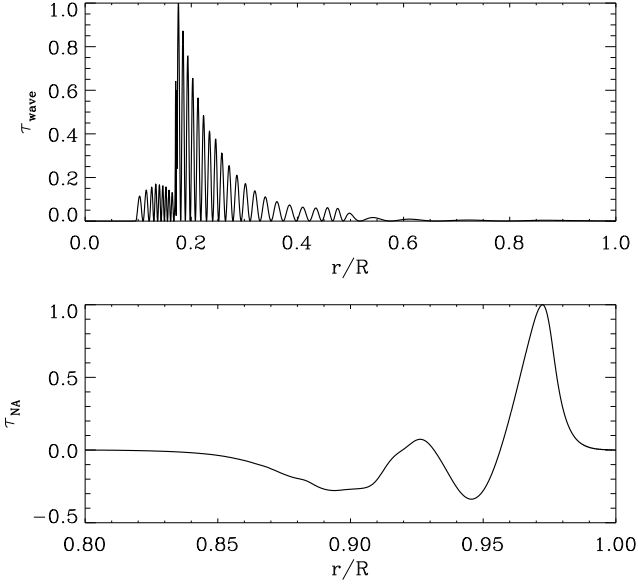}
\caption{$\tau_{\rm wave}$ and $\tau_{\rm NA}$ normalized to the maximum value for a $\ell=4$ $g$-mode with
$\nu=436.0\,\mu$Hz. For $\tau_{\rm NA}$ only the external layers are shown since it is negligible at
deeper ones. Model 1 has been used.}
\label{fig:tauna}
\end{figure}

Fig. \ref{fig:tauna} shows $\tau_{\rm wave}$ and $\tau_{\rm NA}$ as given by Eq.~\ref{eq_tauwave} and \ref{eq_tauna}
respectively for an $\ell=4$ unstable mode. According to 
Eq. \ref{eq:changej}, these functions give the contribution of each point in the star to the change in the angular 
momentum. The wave transit term, $\tau_{\rm wave}$, is similarly weighted as the integrand in the dimensionless 
energy ${\cal{E}}$ (compare to third panel in Fig.~\ref{fig:autof}), 
while $\tau_{\rm NA}$, even for inner gravity modes, is only important in the uppermost layers where non-
adiabatic effects are large. The relative importance between both terms depends on the energy of the mode and the value 
of $\sigma_{\rm I}$ considered in $\tau_{\rm wave}$ and will be discussed in the next section.

Since sdB stars are slow rotating stars, in principle Eqs.~\ref{eq_tauwave} and \ref{eq_tauna} would be enough for 
estimating the changes in rotation. However,
prograde and retrograde modes contribute in an opposite way to the angular momentum transfer. 
If modes with the same degree but azimuthal orders $\pm m$ had the same amplitude, the net effect would cancel out.
For this reason we will also consider the full expression for $\tau_{\rm NA}$ as given by Eq. 36.16
in Unno et al. (1989):
\begin{eqnarray}
\tau_{\rm NAf} & = & \rho \varpi {\rm Im} \left[ {v}_{\phi}^* \,(\sigma + m \Omega ) v_T \frac{\delta S}{c_p}\right] 
\nonumber \\
& &- m \rho {\rm Re} \left[ \frac{({\vec v}_p \cdot \rho^{-1} \nabla_{\ell} p ) v_T \, \delta S^* /c_p} 
{(\sigma^* + m \Omega )} \right] \; ,
\label{eq_taunafull}
\end{eqnarray}
where $\vec{v}_p$, $v_{\phi}$ and $\delta S$ are linear perturbations in, respectively, the poloidal and azimuthal 
component of the velocity, and the entropy. This equation does not neglect $\Omega$ against $\sigma_{\rm R}$
nor does it assume the Cowling approximation. 
However, we shall still consider the eigenfunctions of a non-rotating equilibrium model.

\subsection{Effect in the rotational splitting}

The rotational splitting, i.e. the perturbation in the frequencies caused by rotation, is given
by \citep[e.g.][Eq. 3.349]{aerts10}
\begin{equation}\label{eq:deltaw}
 \delta \omega_{n\ell m}=m \int_0^R\int_0^\pi K_{n\ell m}(r,\theta)\Omega (r,\theta) dr d\theta
\end{equation}
where the kernel $K_{n\ell m}(r,\theta)$ is given by;
\begin{eqnarray}
K_{n \ell m }& =
& \biggl[ \sin\theta \{ |\xi_r(r)|^2 P_{l}^{m} (cos\theta)^2
\nonumber \\
& & +|\xi_h(r)|^2\biggl[\biggl(\frac{dP_l^{m} }{d\theta}\biggr)^2+
\frac{m^2}{\sin^2\theta}P_l^{m} (\cos\theta)^2
\nonumber \\
 &  & -P_{l}^{m} (cos\theta)^2[\xi_r^*(r)\xi_h(r)+\xi_r(r)\xi_h^*(r)]
\nonumber \\
 & & -2P_{l}^{m} (cos\theta)\frac{dP_l^{m} }{d\theta}
\frac{cos\theta}{\sin\theta}|\xi_h(r)|^2\}\biggr] 
\rho r^2 /I_{nlm}
\end{eqnarray}
and
\begin{equation}
I_{nlm} = \frac{2}{2l+1}\frac{(l+|m|)!}{(l-|m|)!}\int_0^R\biggl[|\xi_r|^2+l(l+1)|\xi_h|^2\biggr]\rho r^2dr
\; .
\end{equation}
Therefore, from Eqs.~\ref{eq:tau} and \ref{eq:deltaw}, we obtain the variation of the 
frequency splitting caused by transfer of angular momentum between eigenmodes and rotation:
\begin{equation}
 \frac{\partial \omega_{n\ell m}}{\partial t} = -
m \int_0^R\int_0^\pi K_{n\ell m}(r,\theta)\frac{\tau_{n'l'm'}}{\rho r^2 \sin^2\theta} dr d\theta 
\label{eq:split}
\end{equation}
where $n$, $\ell$, and $m$ are for the mode for which the rotational splitting is considered 
and $n'$, $\ell'$, $m'$ are for the mode for which the transfer of angular momentum is computed.

\begin{figure}
\centering
\includegraphics{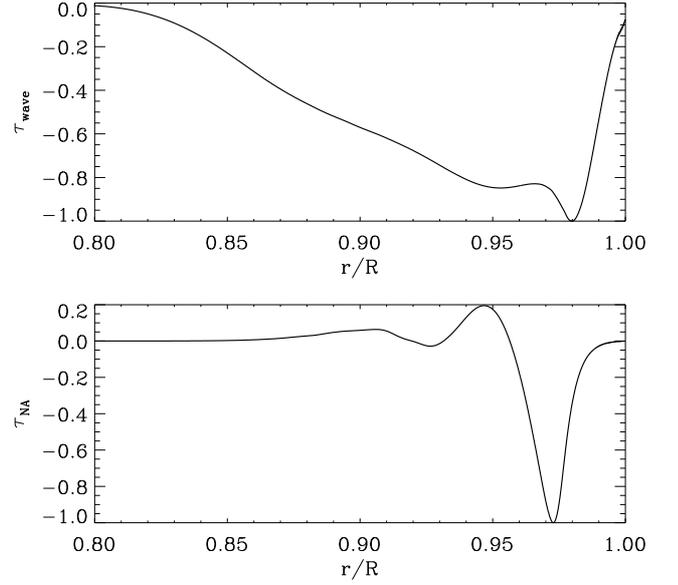}
\caption{Contribution of the different layers to the change in the splitting of the $l=1$, $\nu=2807.5\,\mu$Hz mode
due to the $\ell=4$, $m=4$, $\nu=436\,\mu$Hz $g$-mode according to Eq. \ref{eq:split}. Top panel is for $\tau_{\rm wave}$ 
and bottom panel for $\tau_{\rm NA}$. Values have been normalized to the maximum. Note that only the upper layers of 
the model are shown since the contribution of deeper layers is negligible. Model 1 has been used.}
 \label{fig:intchan}
\end{figure}

Fig. \ref{fig:intchan} shows the normalized outcome of Eq. \ref{eq:split} when either the wave transit term 
$\tau_{\rm wave}$ (top panel)
or the non-adiabatic term $\tau_{\rm NA}$ (bottom panel) are considered. 
Here we have chosen the splitting of the theoretical $\ell=1$ $p$-mode closest in frequency to the observed triplet and 
computed the change due to a single, unstable, $\ell=4$ $g$-mode. 
For $\tau_{\rm wave}$, the change in the splitting is mainly caused in the external layers, 
whereas the angular momentum transfer takes place in deeper regions, see Fig.~\ref{fig:tauna}.
This is a straightforward consequence of the nature of the $p$-mode considered for measuring the rotational splitting
that have larger amplitudes in the external layers.
For $\tau_{\rm NA}$, the external layers are dominant for both the rotational splitting variation and the angular momentum 
transfer, although the weights are different.

\subsection{Visibility}

Since all computations are based on the linear theory, the eigenfunctions are calculated except for a 
constant factor that can be fixed by, for example, the velocity amplitude at a given point in the atmosphere.
In principle velocity observations are easier, less model dependent, and more related to theoretical values than 
photometric observations, and therefore we will use them. 
However, as commented in Sect. 2, the line-of-sight-velocity observations do not correspond to a specific point in the 
atmosphere. Since the amplitude of the
modes change by more than one order of magnitude throughout the atmosphere, an accurate determination would 
require a detailed theoretical simulation of the observational measurements.
In this work we will perform a simple calibration between observed and theoretical amplitudes by assuming 
that the observations correspond to a unique optical depth, $\tau_{\rm R}$, but considering different values.

Specifically, we use a relation between the amplitude of the radial component of the velocity, $v_p$,
and the observed line-of-sight velocity, $v_{\rm obs}$. The proportionality factor, that we will call the 
visibility, $A(\ell,m,\omega)$, is given by
\begin{equation}\label{eq:visibility}
 v_{\rm obs}=A(\ell,m,\omega)\,v_p \quad\mbox{at the observed point.}
\label{eq:vobs}
\end{equation}
Given that we observe $v_{\rm obs}$ for certain modes, we can retrieve their $v_p$ if we are able to compute 
$A(\ell,m,\omega)$. 
The factor $A(\ell,m,\omega)$, includes the geometrical effects of the velocity projection as well as limb darkening effects. 
It also depends on the inclination angle $i$ and the ratio between the horizontal and vertical components of the
eigenfunctions, $K=\xi_h/\xi_r$ at the observed point. 
To avoid the dependence on $i$, the rms value over $m$ can be used. When referring to this average, we will just remove
the $m$ dependence, $A(\ell,\omega)$.

To compute $A(\ell,m,\omega)$ we have mostly followed the formalism in~\cite{aerts92}.
However, in that work the approximation 
\begin{equation}\label{eq:K}
K\approx(GM)/(R^3\sigma^2)
\end{equation}
was used. This approximation is derived by assuming 
a Lagrangian pressure perturbation at the outer boundary of $\delta P=0$, which is valid for adiabatic oscillations,
provided that the boundary condition is not located too deep in the atmosphere. However, 
Eq.~\ref{eq:K} is not valid if non-adiabatic effects are important.
Since we have computed the theoretical eigenfunctions in a non-adiabatic approximation 
for our specific models, we can compute the precise ratio $K$ at the required point in the atmosphere. 

\begin{figure}
\centering
\includegraphics[]{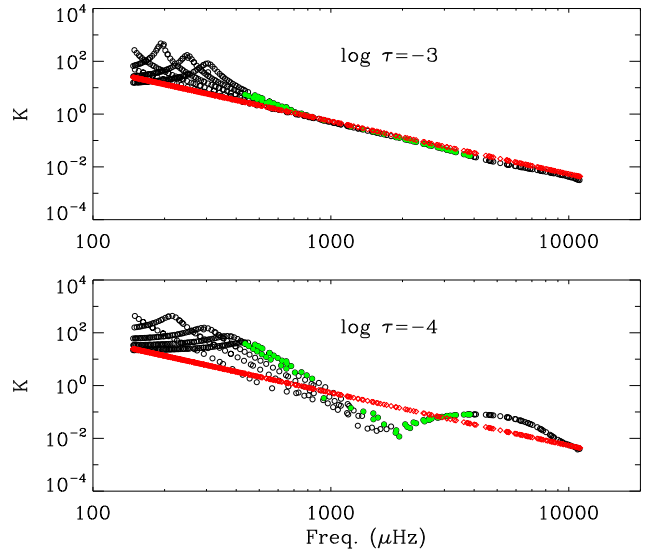}
\caption{$K$ ratio between horizontal and radial components of the displacement vector at optical depths 
$\log\tau_{\rm R}=-3$ and $\log\tau_{\rm R}=-4$. 
Black points include all the linearly stable modes while green points are for the unstable ones (Model 1). 
Red crosses are the ratios expected by the approximate 
Eq.~\ref{eq:K}.}
\label{fig:k_ratio}
\end{figure}

In Fig.~\ref{fig:k_ratio} we show the $K$ ratio evaluated at two optical depths. The approximate $K$ 
from Eq.~\ref{eq:K} is also included for comparison. 
Note that the approximation is in better agreement with the actual theoretical values at higher depths, 
where non-adiabatic effects are less important.
The visibility factors $A(\ell,\omega)$ depend on the optical depth through limb-darkening coefficients and the ratio $K$
which is negligible for $p$-modes. However for $g$-modes it can be as large as 100 or 1000 (as can be seen in 
Fig. \ref{fig:k_ratio}) and as a result $g$-modes can have rather different visibility factors depending on the optical 
depth considered. 

Since the horizontal components of the $g$-modes are very large, 
the visibility factors as defined by Eq.~\ref{eq:vobs} are not by themselves 
indicative of the kind of modes we expect to 
observe. It is more illustrative to express the visibility in terms of the energy of the modes. 
The kinetic energy, averaged over time, can be expressed as \citep[see e.g.][]{aerts10}:
\begin{equation}\label{eq:ekin}
E_{\rm kin} = \frac{1}{2} {\cal{E}} M V_{rms}^2 =
\frac{1}{2} {\cal{E}} M
v^2_p(R) \left( 1+ \frac{ \ell (\ell+1) \xi_h^2(R)}{\xi_r^2(R)} \right)
\end{equation}

\noindent where $\cal{E}$ is given by Eq.~\ref{eq:aener}, $V_{\rm rms}$ is the root mean squared velocity over the 
stellar surface, at radius $R$ and $M$ is the total mass of the star. 
Using Eqs.~\ref{eq:ekin} and \ref{eq:visibility} we obtain the following relation between the line-of-sight velocity 
and the kinetic energy of the modes:

\begin{equation}\label{eq:vobs_ekin}
\frac{v^2_{\rm obs}}{E_{\rm kin}} = \frac{A^2(\ell,m,\omega)}{M\cal{E}} 
\frac{2}{1+ K^2\ell (\ell+1)}
\; .
\label{fig:vobs_ekin}
\end{equation}

\begin{figure}
\centering
\includegraphics{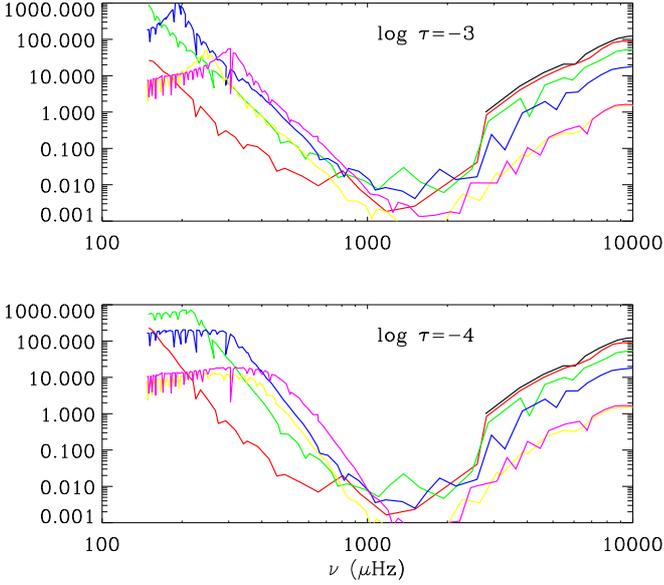}
\caption{Ratio $|v_{\rm obs}|/\sqrt{E_{\rm kin}}$ normalized to the fundamental mode. 
See text and Eq.~\ref{fig:vobs_ekin} for
an explanation. Modes with the same degree have been joined for clarity. Black, red, green,
blue, yellow and magenta correspond to angular degrees $\ell=0$, 1, 2, 3, 4, and 5, respectively.
Upper panel is for $\log\tau_{\rm R}=-3$ and bottom panel is for $\log\tau_{\rm R}=-4$.}
\label{fig8}
\end{figure}

The right hand side of Eq.~\ref{eq:vobs_ekin} can be theoretically computed, 
and it is shown in Fig.~\ref{fig8} for two different optical depths. They have been normalized to the fundamental mode.
Note that $g$-modes in the observed frequency range have much larger values than the fundamental one and hence need 
much less excitation energy to be observed with similar Doppler velocities. This is true not only for the lowest
degrees but also for $\ell=4$ and $\ell=5$. Moreover this conclusion is
valid for any optical depth in the range considered. 

\subsection{Amplitude calibration\label{sec:cal}}

According to Eqs. \ref{eq:taus}, \ref{eq_tauwave}, \ref{eq_tauna} and \ref{eq:split} the transfer of angular 
momentum and the change in the splittings are proportional to the eigenfrequency squares and hence the amplitude square 
of the $g$-modes considered. Since the amplitude of the modes can increase by an order of magnitude throughout the 
atmosphere, a very important issue concerning Eq.~\ref{eq:vobs} is where to fix the point where observed
velocities are transformed to theoretical ones.
Following \cite{rauch10}, the formation depth of the core of the Balmer lines in the subdwarf stars
seems to be at optical depths between $\tau_{\rm R} \simeq 0.1$ and $\tau_{\rm R} \gtrsim 0.01$.
However the wings of the lines are formed higher in the atmosphere and since the amplitude 
of the modes increase with height, the oscillatory signal could come from optical depths
substantially lower than that of the core. 
Here we shall try not to overestimate the change in the angular momentum due to the oscillations. This requires
to assume as small as possible $g$-mode energies, and hence to locate the signal as high as possible in the atmosphere.
We will take $\tau_{\rm R}=0.001$ as a reference value.

Since only few $g$-modes are observed in velocity and they cannot be identified 
($l$ and $m$ are unknown), we need some hypothesis about the excitation energy to be able to proceed.
To calibrate the theoretical amplitudes we attribute a line-of-sight velocity of $v_{\rm obs}=1\,$km/s 
to the $g$-mode with the largest $v_{\rm obs}^2/E_{\rm kin}$ in the frequency range $[250,400]\,\mu$Hz.
The value of $v_{\rm obs}$ and the frequency range considered are based on the observations.
Our choice gives the minimum energy required to reproduce the observational amplitudes in the 
$g$-mode region.
At the two optical depths considered in Fig.~\ref{fig:k_ratio}, the theoretical mode that matches this requirement is a 
$\ell=3$, $\nu=276\,\mu$Hz eigenmode for Model 1. but there are other $l=2,\ldots,5$ modes that if taken as representative 
of the observations would result in a similar $E_{\rm kin}$, so the particular mode is not important for an order of 
magnitude estimation. 

On the other hand, the optical depth of the model at which the line-of-sight velocity is calibrated has a large impact 
on the results. For the above $g$-mode, the amplitude of the radial velocity eigenfunction at an optical depth of 
$\log\tau_{\rm R}=-3$ is 5.3 times smaller than that at $\log\tau_{\rm R}=-4$. Since the change
in the splitting is proportional to the velocity square, a factor of 28 difference is found between both optical depths.

\section{Results}

In the following computations we assume the same kinetic energy for all modes, although there is no particular theoretical 
reason for this. 
However, from the velocity observations, we known that at least a few modes in the frequency range $[250,400]\,\mu$Hz 
need to have such a large $E_{\rm kin}$ and the observations in photometry reveal that in fact a bunch of $g$-modes are 
excited to similar amplitudes and hence, perhaps, energies. 
On the other hand, if $E_{\rm kin}$ varies in orders of magnitude with frequency or angular degree, the actual 
behaviour of the changes 
in the rotational splittings, proportional to $E_{\rm kin}$, can be very different from those shown in the following 
figures. Hence, we can estimate the order of magnitude of the change from individual modes but we are not able to compute 
a net effect from all the $g$-modes.

\subsection{The non-adiabatic term, $\tau_{\rm NA}$}

\begin{figure}
\centering
\includegraphics{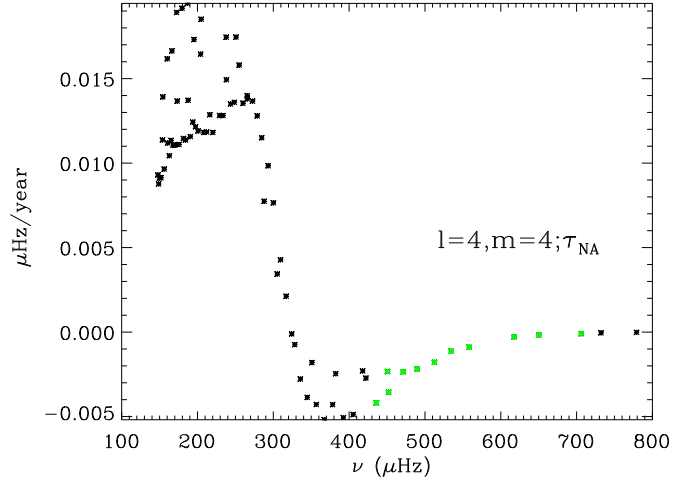}
\caption{Change in the rotational splitting
of the $\ell=1$, $\nu=2807.5\,\mu$Hz $p$-mode due to the $\ell=4$, $m=4$ $g$-modes, considering the $\tau_{\rm NA}$ term. 
Here we assume the same energy for all the modes and fix it so that 
$v_{\rm obs}=1\,$km/s for $\ell=3, \nu=276\,\mu$Hz 
at $\log\tau_{\rm R}=-3$. Black points are for stable modes and green 
points for the unstable ones. Model 1 was used.}
\label{fig2_1}
\end{figure}

Let us start by considering the changes in the splitting due to the non-adiabatic term $\tau_{\rm NA}$ 
of individual $g$-modes with $m>0$.
In Fig. \ref{fig2_1}  we show the change in the splitting of the $\ell=1$,
$\nu=2807\,\mu$Hz mode due to $g$-modes with $\ell=4$, $m=4$.
Eq.~\ref{eq:tauna} for $\tau_{NA}$ and Model 1 have  been used. In the figure, the energy $E_{\rm kin}$ was fixed by 
taking the value $v_{\rm obs}=1\,$km/s at $\log\tau_{\rm R}=-3$ for the reference mode given in Sect. \ref{sec:cal}.
The modes that induce the largest changes in the splitting are in the observed frequency range,
although most of them are not expected to be excited in Model 1. 
It follows from Fig.~\ref{fig2_1} that the change in the splitting of 
the theoretical counterpart of the observed triplet is as large as $0.02\,\mu$Hz/yr if $\log\tau_{\rm R}=-3$ is assumed, 
considering only the contribution of the $\ell=4$, $m=4$ $g$-modes.
Slightly larger values are obtained if $\ell=5$, $m=5$ $g$-modes are considered. 
These changes for single modes are only an order of magnitude smaller than the observed values.
On the other hand, as stated before, calibrating the observed amplitude at an optical depth of $\log\tau_{\rm R}=-4$ 
results in changes in the rotational splittings of about 28 times smaller. 
Thus the resulting splitting for the same single modes would be about $0.007\,\mu$Hz/yr. 

However the net effect could be much smaller since
modes with the same degree $\ell$ and radial order $n$ but different azimuthal orders $m$ can be expected to be excited 
to very similar amplitudes and, according to Eq. \ref{eq:tauna}, the change in the 
rotation due to $\pm m$ pairs would mostly cancel out. 
Note also that high frequency $g$-modes give a contribution opposite to that of low frequency $g$-modes, as seen in 
Fig. \ref{fig2_1}.

\begin{figure}
\centering
\includegraphics{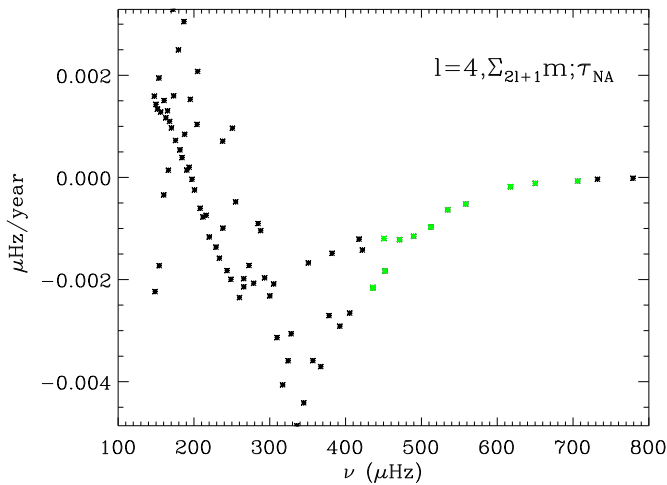}
\caption{Change in the rotational splitting
of the $\ell=1$, $\nu=2807.5\,\mu$Hz $p$-mode due to $\ell=4$ multiplets,
assuming the same amplitude for all the modes with different $m$ but the same $\ell,n$ numbers, and considering the 
$\tau_{\rm NA_f}$ term.
As before we assume the same energy for all the modes and fix it so that the 
$\ell=3, \nu=276\,\mu$Hz mode would have an observed velocity 
$v_{\rm obs}=1\,$km/s if it were to be observed at an optical depth $\log\tau_{\rm R}=-3$.
Model 1 has been used.}
\label{fig13}
\end{figure}

To examine the possible reduction of the net effect, we now assume the same amplitude for all the $2\ell+1$ 
azimuthal modes with the same angular degree $\ell$ and radial order $n$ and use the full expression of $\tau_{\rm NAf}$
(Eq.~\ref{eq_taunafull}) to compute the change in splitting.
Fig.~\ref{fig13} shows the changes in the splitting of the same $\ell=1$, $\nu=2807.5\,\mu$Hz $p$-mode due to the $\ell=4$, 
$g$-modes. The energy of the modes were again fixed by
assuming an optical depth of $\log\tau_{\rm R}=-3$. In this case the change induced by isolated $(2\ell+1)$ group of modes 
can be as large as $0.004\,\mu$Hz/year, which is 60 times smaller than the observed
one. Again, considering the $\ell=5$ modes, slightly larger values are 
obtained while for $\log\tau_{\rm R}=-4$ the effect will be reduced by a factor 28. 

We could sum up the contribution of all the modes thus probably making the change in rotational 
splitting even larger than we give here. However, as explained at the beginning of Sect. 5, 
we do not find it realistic to give a conclusive 
summed up number. Instead, our results should be considered as a low order of magnitude estimation of the change in 
rotational splitting due to angular momentum transfer. 

\subsection{The transient term, $\tau_{\rm wave}$}

\begin{figure}
\centering
\includegraphics{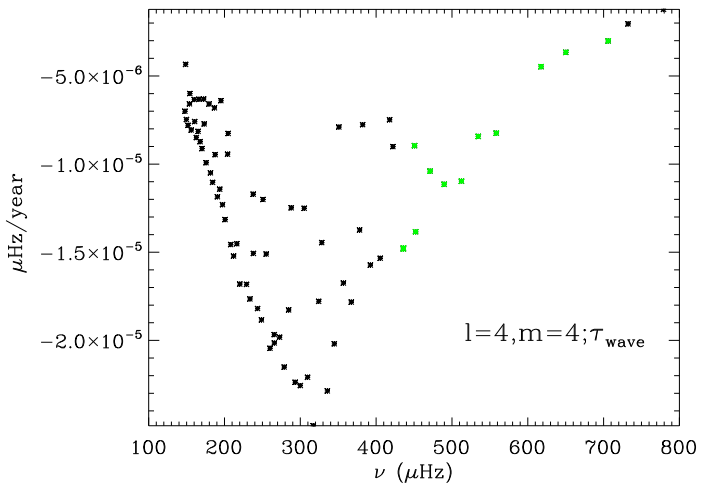}
\caption{Change in the rotational splitting
of the $\ell=1$, $\nu=2807.5\,\mu$Hz $p$-mode due to the $\ell=4$, $m=4$ $g$-modes considering the $\tau_{\rm wave}$ term. 
Here we assume the same energy for all the modes and fix it so that 
$v_{\rm obs}=1\,$km/s for $\ell=3$, $\nu=276\,\mu$Hz at $\log\tau_{\rm R}=-3$. 
Black points are for stable modes and green points for the unstable ones. Model 1 has been used.}
\label{fig2_2}
\end{figure}

Here we consider the changes due to the term $\tau_{\rm wave}$. This term is caused by a change in the mode amplitude,
represented in Eq. \ref{eq:tauwave} by the term $\sigma_{\rm I}$. 
Since the eigenfrequencies are computed in the linear theory, considering the theoretical values $\sigma_{\rm I}$ has 
little sense. On the other hand, as shown in Table \ref{tbl:baran08}, the mode amplitude is changing yearly between
campaigns. These changes can be caused by different factors, such as beating phenomena, but as an upper limit we
can assume they are completely due to intrinsic changes in the mode amplitudes.
A value of $\sigma_{\rm I} = 1/\mbox{year}$ is consistent with the results in Table \ref{tbl:baran08}.
In any case, since the change in the splittings are proportional to $\sigma_{\rm I}$ the results are easly 
transformed to other amplitude variations. Figure \ref{fig2_2} shows the results in the splitting of the 
$\ell=1$, $\nu=2807.5\,\mu$Hz $p$-mode due to the $\ell=4$, $m=4$ $g$-modes. Compared to Fig. \ref{fig2_1} we see
that this changes are about 100 times smaller than those corresponding to $\tau_{\rm NA}$. 
Thus, unless the changes in the mode amplitudes have time scales much shorter than a year, 
the contribution of $\tau_{\rm wave}$ can be neglected as compared to the $\tau_{\rm NA}$ effect. 

\subsection{Changes in the splitting for different modes}

\begin{figure}
\centering
\includegraphics{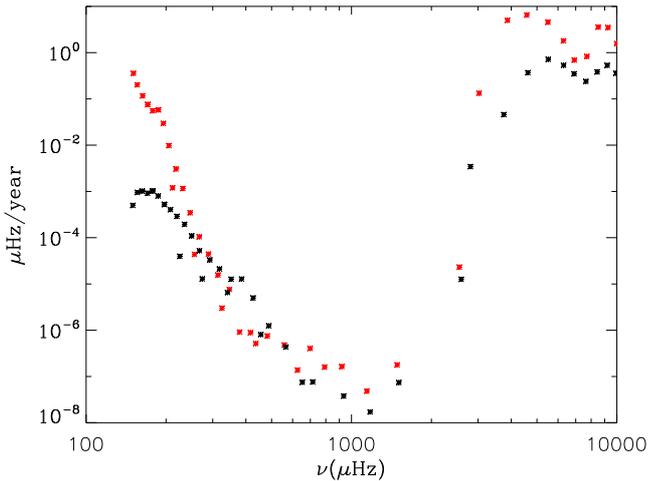}
\caption{Change in the rotational splittings of the $\ell=1, m=1$ modes
due to the and $\ell=4$, $m=4$, $\nu\simeq 300\,\mu$Hz mode.
Absolute values have been considered.
The energy of the mode was fixed in such a way that 
$v_{\rm obs}=1\,$km/s for $\ell=3$, $\nu=276\,\mu$Hz at $\log\tau_{\rm R}=-3$ for Model 1.
Black point are for Model 1 and red points for Model 2.
The $g$-mode considered has $\nu=305.14\,\mu$Hz in Model 1 and $\nu=301.02\,\mu$Hz in Model 2.
}
\label{fig2_1b}
\end{figure}

So far we have limited the computations to the changes in the observed triplet (assuming a $\ell=1$, low order
$p$-mode). However, the change in the splitting strongly depends on the frequency of the $p$ or $g$-mode used as the test.
On the other hand, the expected change for the $\ell=2$ modes are very similar to that of the $\ell=1$ when
compared at a given frequency, except for a factor
$m$ for $p$-modes and a factor $m\{1-1/[l(l+1)]\}$ for $g$-modes, both well known results from the asymptotic theory.
To illustrate the frequency dependence, Fig. \ref{fig2_1b} shows
the changes in all the $\ell=1$ modes due to a $g$-mode with $\ell=4, \nu\sim 300\,\mu$Hz.
From Fig. \ref{fig2_1b} follows that the largest changes are 
expected for $p$-modes with high radial order. The observed triplet is around $2800\,\mu$Hz but higher 
$p$-modes are in fact observed. So this result
is potentially very interesting since we predict changes in the splitting around or above $1\,\mu$Hz/year. 
Also, the change in the rotational splitting of low frequency $g$-modes would be significantly larger than those 
corresponding to the observed modes, but measuring rotational splitting in this region could be a difficult task due
to the large population of peaks and beating phenomena in the signal.

Since $\tau_{\rm NA}$ is proportional 
to the work function, the change in the splittings are determined by the same non adiabatic properties of the $g$-mode 
eigenfunctions as the growth rates, although the integrals over the whole star have different weights.
Since the modes with the largest contribution to the changes in the rotational splittings are stable in Model 1, this can 
be an important source of errors. On the other hand, as shown in Fig.~\ref{fig:growth}, $g$-modes with $l\leq 5$ and low 
frequencies, are unstable in Model 2 and hence, in principle, using this new model would avoid that problem. 
Fig. \ref{fig2_1b} allows to analyze this problem.
The $g$-mode chosen for the transfer of angular momentum is expected to be stable if Model 1 is used and unstable if Model 
2 is used. Despite this fact,
Fig. \ref{fig2_1b} shows that the results for both models are rather similar, or a little larger for Model 2,
if compared at the same frequency of the test mode.
This allows us to be confident in the results obtained with Model 1.
The reason for using Model 1 in the previous figures was that the frequencies of the $l=1$ and $l=2$ 
lowest $p$-modes are very close to the observed triplet and quintuplet
while, as previously mentioned, we do not yet have a model with the new physics that accurately match the observed 
$p$-modes.

\section{Conclusions}

Transfer of angular momentum between $g$-modes and rotation can explain the changes in the rotational splittings observed 
in the sdB star Balloon\,090100001. We have done the computations by estimating the mode energies from the 
observed $g$-mode velocities and using structure models that fit the stellar parameters, including the frequencies of the
highest amplitude $p$-modes. The theoretical changes found for the rotational splittings due to single $g$-modes are
about one order of magnitude smaller than the observed ones if an optical depth of $\log \tau_{\rm R}=-3$ is considered
for the calibration. 
If the cancellation effect between prograde and retrograde modes are considered by assuming the same amplitude
for the rotational multiplets, the net effect is about 60 times smaller than the observed ones. 
Since the optical depth considered is rather a lower limit and because the observed value should be compared
with the net effect from all $g$-modes, we can conclude that this mechanism can produce a rotational splitting of the same 
order of magnitude as the observed ones.
Moreover, we find that the transfer of angular momentum is possible due to the non-adiabatic
nature of high radial order, low degree $g$-modes, even if the amplitude of the modes do not vary. In fact, changes in 
the amplitudes within a yearly time scale, as it seems to be the case in this and other sdB stars, induce a much smaller 
change in the rotational splittings than the non-adiabatic process. 

Although Ba09 has brought us an opportunity for checking this kind of stellar phenomena, some uncertainties prevent us for 
obtaining more quantitative results. One uncertainty comes from performing an accurate calibration between observed 
velocities and theoretical energies, because the former cannot be easily translated to values at an specific optical depth 
or to a some weighted integral over the atmosphere. 
Photometric observations can help here, but a calibration of such observations are generally more complex and
model dependent than the velocity ones since changes in the flux are due to non-adiabatic effects. 
This problem could be limited in the future by performing detailed theoretical simulations of the data analysis.

A second problem that hinders more quantitative calculations concerns the mode identification in the 
low frequency range. But even if this were possible the main contribution to the transfer of angular 
momentum can come from high degree modes that are not observed due to geometrical cancellations.
However it is interesting to note that potentially observed modes and modes contributing
the most to the changes in the rotational splitting can be rather close in frequency and degree.
As can be seen in Fig. \ref{fig:growth}, Model 2 predicts unstable modes with a frequency range that resemble the observed 
one. There are unstable modes for all the angular degrees considered, but for higher degrees the unstable modes shift 
towards higher frequencies. Thus the main contribution, that according to Figs. \ref{fig2_1} and \ref{fig13} are for 
frequencies below 400 or $500\,\mu$Hz could come from modes of degrees in the observed range or at most a little higher, 
say $\ell\leq 5$ or 6. 

According to Eq. \ref{eq:tau}, since $\tau$ depends on latitude, so does the change on the rotational velocity. Hence, 
$g$-modes induce a differential rotation in the stars. It is interesting to note that the quintuplet did show a 
latitudinally differential rotation, and hence the same angular momentum transfer can be invoked as the cause for the 
differential rotation.

Searching for other sdB stars with similar characteristics would allow to improve our knowledge.
This involves in particular space mission like COROT and KEPLER \citep{Borucki2010}
that can acquire high precision and long-term photometry of pulsating hot subdwarfs, which is of utmost importance to 
detect the possible effects in frequency splittings due to angular momentum transfer. However, there is only one $p$-mode 
pulsator identified in the FoV of Kepler \citep{kawaler2010} and a handful of $g$-mode pulsating sdBs with candidate 
splitting patterns that still need longer time series to be confirmed \citep{ostensen2010,ostensen2011}.

\begin{acknowledgements}
Part of this work was supported by the Spanish National Research Plan under project AYA2010-17803. 
HH is supported by a Rubicon fellowship of the Netherlands Organisation for Scientific Research (NWO).
\end{acknowledgements}

\bibliographystyle{aa}
\bibliography{biblio}

\end{document}